\documentclass[final,authoryear,5p,times]{elsarticle}

\usepackage{graphicx}
\usepackage{amssymb}
\usepackage[utf8]{inputenc}
\biboptions{}

\journal{Astronomy \& Computing}

\begin{document}

\begin{frontmatter}



\title{Virtual Observatory Publishing with DaCHS}


\author{Markus Demleitner}
\author{Margarida Castro Neves}
\author{Florian Rothmaier}
\author{Joachim Wambsganss}

\address{Unversität Heidelberg, Zentrum für Astronomie,
Astronomisches Rechen-Institut, M\"onchhofstraße 12-14, 69120
Heidelberg, Germany}

\begin{abstract}
The Data Center Helper Suite DaCHS is an integrated publication package
for building VO and Web services, supporting the entire workflow from
ingestion to data mapping to service definition.  It implements all major
data discovery, data access, and registry protocols defined by the VO.
DaCHS in this sense works as glue between data produced by the data
providers and the standard protocols and formats defined by the VO. This
paper discusses central elements of the design of the package and gives
two case studies of how VO protocols are implemented using DaCHS'
concepts.

\end{abstract}


\end{frontmatter}


\section{Introduction}
\label{sec:intro}
To aid in the adoption of Virtual Observatory (VO) standards, it is
important to keep the entry barrier to running interoperable services low.  In
particular for the VO's ``S-protocols'' 
(SCS\footnote{Simple Cone Search, a protocol allowing remote
searches in catalogs of celestial objects \citep{std:SCS}.}, 
SIAP\footnote{Simple Image Access Protocol, a protocol allowing
discovery of images of the sky \citep{std:SIAP}.}, SSAP\footnote{Simple
Spectral Access Protocol, a protocol supporting the discovery of spectra
\citep{std:SSAP}.}), an
important design goal has been a straightforward mapping to common
network programming pa\-ra\-digms.  Nevertheless, running fully compliant
services seamlessly integrated into the Virtual Observatory as a
whole requires a
significant effort -- for example, a registry record needs to be
maintained, and the auxiliary services required by VOSI\footnote{Virtual
Observatory Support Interfaces, a suite of simple endpoints allowing
clients an inspection of a service's data content, access options, and
health \citep{std:VOSI}.}
have to be in place as well.

With advanced protocols like 
TAP\footnote{The Table Access Protocol \citep{std:TAP} enables the exchange
of database queries and results between clients and remote servers.  For
its operation, it depends on several other standards, in particular the
SQL-like query language ADQL \citep{std:ADQL} and the Universal Worker
Service (UWS, \citealt{std:UWS}) pattern for asynchronous execution.} the
implementation effort is significantly larger.
Therefore, packaged standard software that contains all the building
blocks necessary for the operation of services in a mutually compatible
form is an important contribution to keeping the operation of VO
services within reach of modest organisations.

\begin{figure}[t]
\includegraphics[width=\hsize]{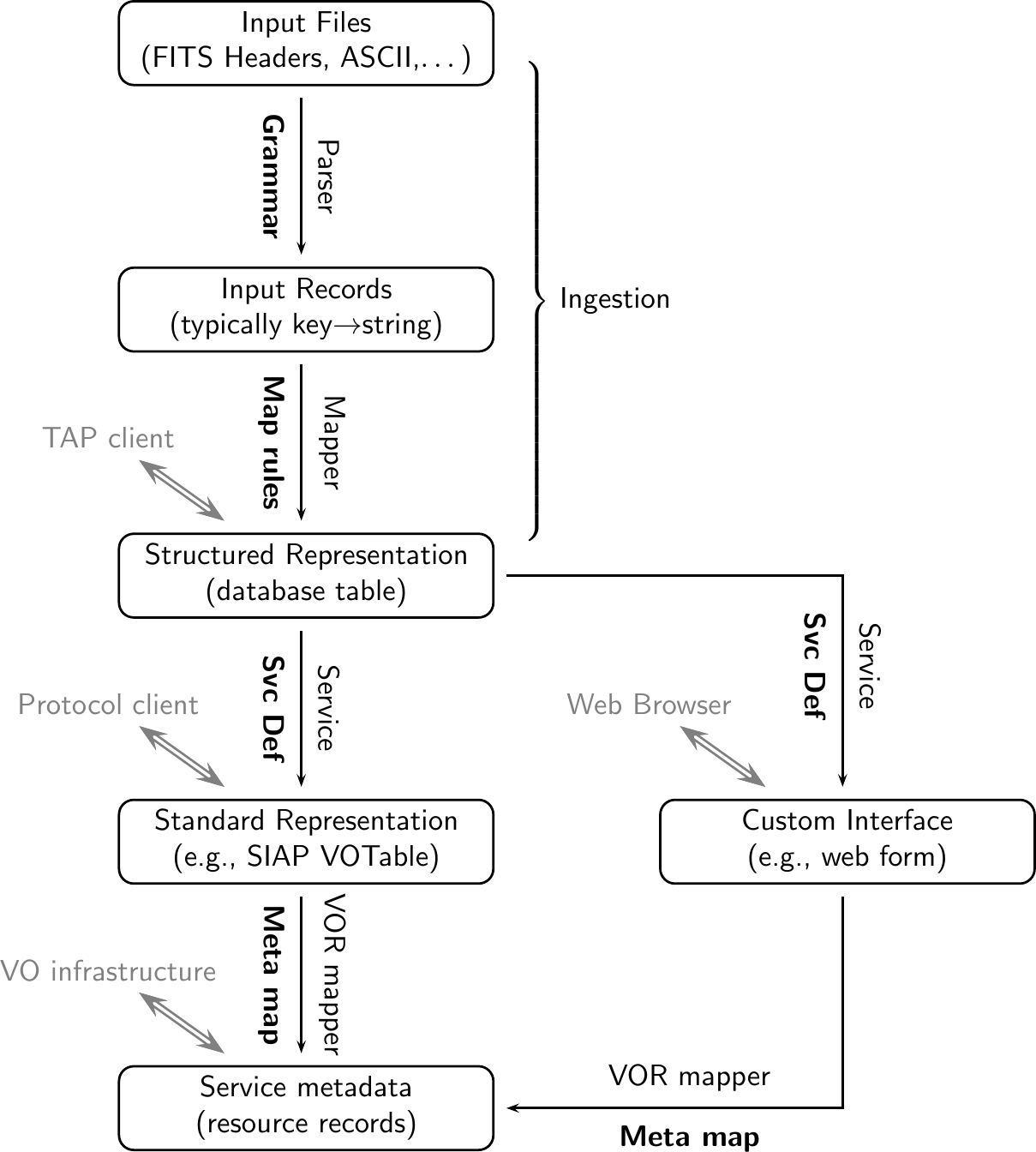}
\caption{The generic workflow leading to a VO publication as modelled by
DaCHS, with boxes
depicting data representations, and single arrows showing the
transformations, annotated with the procedures (above the arrows) and definitions
(boldface, below the arrows) involved in the transformation.  In gray
with double arrows we indicate external users of the various
representations.}
\label{fig:dcop}
\end{figure}

The German Astrophysical Virtual Observatory (GAVO) has been developing such a
package since 2007 under the name DaCHS (which stands for Data Center
Helper Suite).  Compared to most other packages available for this
purpose (for instance, VO-Dance \citep{2012SPIE.8451E..05M}
or Saada; for more information on
packages available for VO publishing, see the IVOA's web page on
publishing in the
VO\footnote{http://wiki.ivoa.net/twiki/bin/view/IVOA/PublishingInTheVONew}), its focus is on a unified handling of the entire publication
process from the raw data files to the dissemination of data and
metadata, including registry records
(schematically shown in Fig.~\ref{fig:dcop}).  DaCHS also helps in
organising ancillary tasks that may be necessary to unify data for
publication purposes, like header normalization or astrometric
calibration as required for effective publication via VO image access
protocols.  However,
actual data reduction tasks are not considered in scope, which means
that DaCHS will not, for example, grow an actual workflow engine, and
the inputs are expected to be at least almost science-ready.

In DaCHS' development, two main principles served as guidelines:

\begin{enumerate}[(1)]
\item Be declarative when reasonably possible
\item There is exactly one place for each piece of metadata
\end{enumerate}

Point (1) means that when making design choices, we have a bias towards
declarations (``My problem is X'') as opposed to procedural definitions
(``if a, do b, otherwise c'').  This is motivated by the expectation
that declarative specifications will allow easier development of the
underlying software without requiring adaptation of the service
definitions. For an operator running hundreds of services, such a
requirement might otherwise block software upgrades indefinitely.
On the other hand, for many data publication tasks a procedural
specification is much more natural, and DaCHS does take advantage of
that. This is why the guideline is qualified by ``when reasonably
possible''.

Point (2) is motivated by the observation that while quality metadata is
paramount to smooth VO operation, metadata maintenance rarely gets the
attention necessary.
Hence, it must be made as easy
and labour-efficient as possible.  Also, various elements of the data
publication and access process share the same
pieces of metadata. For example, the data type of a database column is
relevant during ingestion, for the service operation itself, and to the
VOR mapper\footnote{VOR or VOResource is used here to refer to the
entire data model of the VO registry as defined through the VO resource
specification \citep{std:VOR} and its extensions.} 
in the model of Fig.~\ref{fig:dcop}. Similarly, a datum like
the bibliographical source is required both in running the service -- at
least when following the  recommendation from DALI\footnote{Data Access
Layer Interface, a set of rules and recommendations common to all
(future) VO data access protocols \citep{std:DALI}.} to include
it in result tables -- and the VOR mapper (i.e., the component creating
registry records; see below).

This tight coupling of the publication workflow through the metadata was
the main driver for producing an integrated publication package,
allowing a re-use of data descriptions all the way from parsing 
the data files to the
generation of registry records.  Weighing this up against the obvious
advantage that a more componentised architecture would have with respect to
easy integration into existing software landscapes, we decided for
an integrated package.

As a publication package brings together several different parties, some
nomenclature must be introduced.  Where we speak of authors, we refer to the
creators of DaCHS itself.  The operators are the persons running the
software, i.e., publishing data.  Users (humans) and clients (software
programs) finally are the consumers of the data.

The term ``publication'' is also ambiguous in this context.  It here refers
to both the process of making some data collection available via network
protocols, and the explicit step of making a resource known to the VO
Registry.  Where the distinction needs to be made, we write
``publication to the Registry'' if we mean this second reading.  
A publication in the sense of ``journal article'' is referred to as
bibliographical source here.

In the remainder of this paper, we will describe DaCHS' view of the
publishing workflow in more detail before dis\-cus\-sing the various
metadata systems in section~\ref{sect:metadata}. 
We then move on to two case studies intended to
illustrate how VO protocols sit on top of DaCHS' infrastructure.  The
first of these (section~\ref{sect:ssap}) 
covers the Simple Spectral Access Protocol SSAP as the
most advanced of the currently defined ``S-protocols'', the second 
(section~\ref{sect:datalink}) the
new datalink protocol that significantly departs from the usual model of
basically providing a network layer on top of a database.

\section{The Publication Workflow in DaCHS}

The publication workflow as modeled by DaCHS is shown in 
Fig.~\ref{fig:dcop}.  It gives the operations necessary to transform a
set of inputs (e.g., images, spectra, tables)
as delivered by the data provider (e.g., research group, instrument
operator)
to the structured
representations required by various consumers -- protocol clients, web
browsers, and the VO Registry.

In explaining the workflow, we illustrate the abstract steps with
samples of inputs DaCHS operators would have to produce.  This
exposition is based on material from \citet{demleitner2013multiprotocol} and
in particular the DaCHS tutorial \citep{soft:Dachstut}.  This latter
resource should be consulted by new operators attempting to reproduce
the steps sketched out here.

In DaCHS, service creation involves writing a \emph{resource
descriptor} (which is discussed in a more general setting in
section~\ref{sect:rd}; we sometimes write RD for short). 
This is an XML file allowing the software to
perform the various tasks shown in \ref{fig:dcop}.  When writing one,
one typically starts with the description of the structured
representation (the center box in Fig.~\ref{fig:dcop}), which in DaCHS
consists of metadata for database tables.  A fragment of such a table
definition could look like this:

\begin{verbatim}
<table id="main" onDisk="True" adql="True" 
  mixin="//scs#q3cindex" primary="hipno">
  <stc>
    Position ICRS Epoch J2000.0 
      "raj2000" "dej2000" 
      Error "err_ra" "err_de"...</stc>
  <column name="hipno" type="integer" 
    ucd="meta.id;meta.main"
    description="Number of the star in ..."/>
  <column name="raj2000" 
    type="double precision"
    ucd="pos.eq.ra;meta.main" unit="deg"...
\end{verbatim}

Glossing over the details, the operator here gives various properties of
the table itself (name, primary key, access control), followed by a
specification of the structure of space-time coordinates in the table
(in this case, this is just a spherical position with errors; see also
the STC discussion in section~\ref{sect:metadata}).  Subsequently, column
metadata -- names, types, units, UCDs\footnote{Unified Content
Descriptors, a controlled vocabulary for expressing terms important for
expressing astronomical metadata like ``pos.eq.ra'' or ``meta.ref.url''
\citep{std:UCD}.} -- is given, one element per column.

Finally, the \texttt{mixin} attribute in the above fragment deserves
more attention.  In
DaCHS, a \emph{mixin} essentially is a collection of aspects of data
representation, first and foremost specific columns, but possibly also
pieces of metadata, indices, and the like.  Additional behaviour
(such as filling the table of file-like products) can optionally be
attached to them.  Mixins are defined in (typically built-in) resource
descriptors and referenced by their identifiers, which in DaCHS consist
of the resource descriptor name, a hash, and an XML id.  The double
slash in front of the resource descriptor name indicates that a
built-in resource descriptor is referenced; these come with the
software distribution and contain -- potentially operator-customiseable
-- material for re-use in operator RDs, as well as descriptions of system
services.

The mixins guarantee certain properties in the table that
protocols exposing the contained data require.  In the example, the
mixin ensures that there is one designated spherical position
(identified via special UCDs) per table row, and that positional queries
over it are fast thanks to Q3C-based indexing \citep{soft:q3c}.  This is
what DaCHS' support for the IVOA SCS protocol builds on.

Another example for a mixin linked to an IVOA protocol is \texttt{//siap\#pgs} 
which endows a table with the columns required in responses for 
SIAP, plus further columns allowing efficient
queries over such an image collection using the pgSphere postgres
extension.  Similar mixins exist for the spectral access protocol,
Obscore\footnote{Obscore is a data model with a a specific
table schema for observational products; coupled with TAP, it allows
complex queries to be run against standardised descriptions of
observational products \citep{std:OBSCORE}.} conformance, and more.

After the definition of the internal representation, the operator next
describes the ingestion process, i.e., whatever is necessary to bring
the input data as provided by the producer to the internal
representation.  In DaCHS, ingestion typically is a two-step process.
In the first step (corresponding to the topmost arrow in
Fig.~\ref{fig:dcop}, a parser produces a sequence of mappings
(``associative arrays'') from the input files, where usually both keys
and values are simple, flat strings.  The descriptions defining how a
given input file relates to the sequence of mappings in DaCHS are
collectively known as grammars, where different types of inputs (FITS
files vs.~CSV, say) require different sorts of rules.  The resulting
mappings are called rawdicts (``raw dictionaries'') in DaCHS
terminology.

For example, if the input comes as a formatted ASCII table, the grammar
would assign labels to column ranges like this:

\begin{verbatim}
<columnGrammar topIgnoredLines="9">
  <colDefs>
    hipno:     3-8
    srcSel:    47-49
      ...
\end{verbatim}

-- which instructs the parser for column grammars to ignore the first
nine lines, and then, in each line, use the contents of columns 3
through 8 to get the value for the key hipno, analogously construct the
value for srcSel, and so forth.

For comparison, when an input format already is highly structured, as in
the case of FITS headers, the grammar specification can be as simple as

\begin{verbatim}
<fitsProdGrammar/>
\end{verbatim}

This tells DaCHS to fairly directly use pyfits
\citep{1999ASPC..172..483B} to create one rawdict each from (in this
case) the primary header of each input file; no additional
instructions are necessary in the simplest cases, as the parsing rules
for FITS headers are already well-defined by an external standard.

In addition to various sorts of text-based parsers, DaCHS also has
parsers built in for FITS binary tables (one rawdict per row), VOTables,
or generic binary records.  Operators can furthermore write custom
parsers in Python.

To actually feed a database table, a second step in the ingestion is
necessary, in which the rawdicts are processed to data structures mapping
column names to typed (and possibly digested) values.  Within DaCHS,
these  are called rowdicts.  

The transformation of rawdicts to rowdicts (the second arrow in
Fig.~\ref{fig:dcop}) is performed by
data mappers (``rowmakers'').  This usually involves more than type
conversion and key mappings: unit conversions, combining inputs (e.g.,
date-time values), arbitrary mapping of values (e.g., standardization of
object or filter names), detection of NULLs,
or computing new values (e.g., waveband limits
from filters) are just some of the tasks regularly necessary to ensure
compliance to IVOA standards or rationalise data representation.

To support this wide variety of mapping operations, 
we allow several levels of mixing in procedural content: plain Python
expressions as mapping values, applying pre-defined 
``procedures'', usually passing parameters, 
or writing such procedures from scratch.  Writing
complete procedures from scratch obviously is the most expressive
method, but as such procedures usually reference implementation details,
it also entails the lar\-gest potential for incompatibilities as the core
software develops.  It is much easier for the core software to keep the
interface of pre-defined procedures and the namespace visible to the
plain Python expression stable.

For instance, a simple rowmaker could look like this:

\begin{verbatim}
<rowmaker idmaps="*">
  <map key="src_sel" source="srcSel"/>
  <map key="raj2000">hmsToDeg(
    @alphaHMS, None)</var>
</rowmaker>
\end{verbatim}

-- this would instruct DaCHS to produce values for \texttt{src\_sel} in
the rowdicts from the values of \texttt{srcSel} in the incoming rawdicts,
converting types as necessary according to default rules, and to use the
built-in \texttt{hmsToDeg} function to parse the values of
\texttt{alphaHMS} in the rawdicts to obtain the values for
\texttt{raj2000} in the rowdicts.  The \texttt{idmaps} attribute finally
says that all remaining keys from the rawdict that have
identically-named counterpart in the database table are to be
type-converted using default rules.

Table mixins are typically accompanied by rowmaker procedures aiding in
the mapping from rawdicts to the specific columns provided by the
mixins.  This is an important mechanism for elementary input validation
and the provision of defaults.  For instance, image metadata necessary
for publication over SIAP, together with some elementary manipulation of
header values, could be covered by a rowmaker like (@key is a shortcut
notation to access values from the rawdict)

\begin{verbatim}
<rowmaker>
  <var name="cleanedObject">
    @OBJECT.split("_")[0]</var>
  <apply procDef="//siap#setMeta">
    <bind key="title"
      >@cleanedObject+" "+@DATE_OBS</bind>
    <bind key="instrument"
      >"%s %s"%(@TELESCOP, @OBSERVAT)</bind>
    <bind key="dateObs">@DATE_OBS</bind>
\end{verbatim}

The \texttt{apply} element is a container for procedural python code
used in rowdict generation. In this instance a predefined procedure
(selected, as usual, by a reference into the built-in \texttt{//siap}
resource descriptor) is invoked with parameters defined in
\texttt{bind} elements.

Grammar, rowmaker, and a specification of what files to read from
(typcially a shell pattern with optional blacklisting) together form an
ingestion rule, which in DaCHS takes a form like

\begin{verbatim}
<data id="import">
  <sources>data/*.txt</sources>
  <columnGrammar topIgnoredLines="9">
    ...
  <make table="main">
    <rowmaker>
      ...
\end{verbatim}

With this, DaCHS has enough information to create and populate the
database table.  This is effected by executing a command like

\begin{verbatim}
gavo imp q
\end{verbatim}

Here, \texttt{gavo} is the name of the DaCHS executable (which is not
called something like ``dachs'' by default for historical reasons). 
The \texttt{imp} argument selects
the import subcommand, which looks for all \texttt{data} elements in the
resource descriptor referenced by its argument and executes the
ingestions described there. Finally, \texttt{q} is the name of the file
containing the resource descriptor -- as the canonical extension for
resource descriptor files is ``rd'', this command line would try evaluating
a file \texttt{q.rd}.

Ideally, this
ingestion would only be done once at service creation time.  In practice, bugs
in the input data, the grammar, or the mapping rules, evolution of the
data collection itself, or improvements in the data publication
routinely necessitate re-ingestions, possibly years after a resource has
been published.  This is the main reason for keeping the ingestion rules
in the resource descriptor.  Keeping them
close to the service and metadata definitions has also proven useful
simply for documenting (executably, if need be) what operations have
been performed during ingestion.

Even if ingestion typically is a rare event, 
for data producing more than a few million
rows it still may become a bottleneck, as the described
operations typically only process a few thousand rows per second on
current hardware.  This is partly due to both the parser and the mapper
being compiled into Python bytecode rather than native object code,
partly due to the overhead of going through SQL serialization and
deserialization of multi-INSERT statements.

To nevertheless allow rapid ingestion of datasets in the gigarecord
range, DaCHS supports a shortcut mechanism called ``direct grammar''.  
These are external binaries
creating material going into the database via binary copy in one step.
Thus, they sidestep DaCHS' mapping mechanisms and combine the roles of
the normal grammars and the rowmakers in one piece of code.  The usual
way of creating such external binaries is to use DaCHS' \texttt{mkboost}
subcommand to generate C
source code templates.  When parsing from
FITS binary tables, the generated source code will immediately work if
the relation between database table and source table is sufficiently
simple.  In all other cases, manual translation of grammar and mapping
rules to C code is required.  The reward of the additional effort and
the loss of declarativeness is that the ingestion time typically is
significantly shorter than the time the database engine spends on
indexing and other post-ingestion operations.

The structured representation resulting from the ingestion is the basis
for the operation of services.  Services adhere to certain protocols
that govern the actual bytestreams of the inputs and outputs of the
service.  That is true regardless of whether the client is a web browser
or specialised software speaking IVOA protocols.
Most of the protocol logic is hardcoded in so-called
\emph{renderers}, which are objects that convert between whatever is on
the wire -- parameters and uploads on input, serialised byte streams on
output -- and internal representations -- input tables on input, output
tables on output.  In DaCHS, they are referenced in short strings
(``form'' for exposing a service over HTML forms, ``siap.xml'' for
handling requests using IVOA's SIAP protocol, ``availability'' to return
VOSI availability information on the service, and so forth).

The actual functionality, based on the internal
representations and performing the computations or queries necessary to
fulfill the incoming query, is provided by \emph{cores}; the most
common one is the dbCore, which generates a database query from incoming
parameters.  For example, here is a simple service definition giving
both a web form and an IVOA cone search service on the table sketched
above, allowing an additional constraint on a column named \texttt{mv}.

\begin{verbatim}
<service id="cone" allowed="scs.xml,form">
  <dbCore queriedTable="main">
    <FEED source="//scs#coreDescs"/>
    <condDesc buildFrom="mv"/>
  </dbCore>
  <outputTable verbLevel="20"/>
</service>
\end{verbatim}

To actually run the services defined in this way, DaCHS has a built-in
server component that, during development, is usually operated in its
debug mode by executing

\begin{verbatim}
gavo serve debug
\end{verbatim}

The services can now be used, with the access URLs derived
from file system paths to the RDs, XML ids within them and renderer
names using a simple scheme.  To make the services
discoverable, however, a publication step is necessary, either locally to the
portal page or globally to the VO Registry.  A sensible publication also
needs carefully written and comprehensive metadata.  Within a resource
descriptor, metadata comes in dedicated metadata elements, typcially at
the head of the resource descriptor, for instance:

\begin{verbatim}
<resource schema="arihip">
  <meta name="title"
    >ARIHIP astrometric catalogue</meta>
  <meta name="description">
    The catalogue ARIHIP has been...
  </meta>
  <meta name="creator.name"
    >Wielen, R....</meta>
  <meta name="subject">Catalogs</meta>...
  <meta name="coverage">
    <meta name="profile">AllSky ICRS</meta>
    <meta name="waveband">Optical</meta>
  </meta>
  <meta name="_longdoc" format="rst">
    The ARIHIP Catalogue is a suitable...
  </meta>
  <meta name="source">2001VeARI..40....1W</meta>
\end{verbatim}

For a full publication to the VO, a VOResource
\citep{std:VOR} record needs to be created. This is an XML fragment
conforming to a set of
XML schema files and collecting a comprehensive and standard set of
service metadata.  Once the metadata is specified, DaCHS creates those
automatically, where capabilities and interfaces declared depend on the
renderer used -- for instance, the \texttt{form} renderer yields
capabilities with interfaces of type \texttt{vr:WebBrowser}, while the
\texttt{ssap.xml} renderer capabilities of type
\texttt{ssap:Sim\-ple\-SpectralAccess}.  All this is largely transparent
to the operator.  

Since publications should not happen accidentally,
they are two-step processes, in which first one or more \texttt{publish}
elements are added in the service element as shown above:

\begin{verbatim}
<publish render="scs.xml" sets="ivo_managed"/>
<publish render="form" 
  sets="ivo_managed,local"/>
\end{verbatim}

-- meaning that the form-based browser interface is advertised both on
the portal (``local'') and to the VO as a whole (``ivo\_managed''),
whereas the cone search interface that requires a specialised client can
only be located through the VO Registry.

After that, 

\begin{verbatim}
gavo pub arihip/q
\end{verbatim}

actually adds the services and data collections
marked for publication with the resource descriptor
to the set of resources reported to the Registry (where \texttt{arihip}
here is an operator-chosen path component) or on the portal page.
This extra
step is non-trivial in that it tests the completeness and, in part,
formal correctness of the metadata that goes into the resource record.

The VOResource records generated by DaCHS need to become part of the VO
Registry at this point.  This happens when a searchable registry
harvests it as discussed in \citet{acregistry}.  
To facilitate this harvesting, DaCHS supports the OAI-PMH
\citep{std:OAIPMH} protocol employed by VO registries and thus lets
operators run a publishing registry.  Once a DaCHS instance is public
and some server-global metadata is specified, it can self-register at
the registry of registries \citep{std:RofR}, after which no further
operator intervention is necessary for the dissemination of new or
updated registry records.

\section{The Metadata Model}
\label{sect:metadata}

Metadata is the central concept in DaCHS.  However, due to a combination
of considerations involving re-use of concepts presumably already known
to the operator, appropriateness of representation, and ease of
implementation, there are several different sources of metadata within
DaCHS.

First, there is the usual column metadata like name, type, UCD,
description, and the like.  Here, DaCHS largely follows the model of
VOTable's FIELD and PARAM elements, adding some attributes as necessary
(e.g., suggested table heading, relative importance, hints for value
formatting, tags of pertaining table notes).

Second, there is metadata on the space-time coordinates
(STC\footnote{The IVOA has a data model for expressing space-time
coordinates, their derivatives, and ancillary metadata; see
\citet{std:STC}.}).  This includes information on the frame
of sets of coordinates within tables as well as the roles
(``declination'', ``error in proper motion in declination'') played by
the columns.  DaCHS employs a slightly enhanced version of
STC-S\footnote{STC-S \citep{std:STCS}, the ``S'' standing for
``string'', is a technique in which space-time coordinates organised
according to IVOA's STC data model are written as flat strings.}
for their specification.
The introduction of a special handling for this kind
of metadata may be vindicated by pointing out that STC-S is also
used to convey space-time coverage in RDs and that a
key-value representation would be too tedious for human input. 

It should be noted that role assignment in other DaCHS-supported VO data
models, in particular the spectral data model \citep[SDM; ][]{std:SDM}, so far
relies on the utype\footnote{VOTable's utype attribute was introduced to
allow linking entities from data models to elements with VOTable.  See
\citet{note:utypeusage} for a discussion of current practices around utypes.}
attributes within column metadata.  In the context
of ongoing attempts to rationalize data model handling in the VO,
we expect to develop a unified method to express
relations between data models and tables as well as additional
model-specific metadata in the future, which should leave STC as less of
a special case.  We expect, however, to keep special handling for this
kind of metadata.  This is mainly because of its highly structured
nature, the prolific occurrence of references, and the tight coupling
to individual pieces of data.

The third and most interesting class of metadata is the ``open''
metadata.  It is used to hold most of the VOResource metadata like
title, author, technical contact, related resources or test queries
\citep{std:RM,std:VOR}, as well as locally-used metadata (like detailed
documentation, usage hints, or table notes).  

Several objects can hold metadata in DaCHS: the whole data
center, the resource descriptor, tables, services, and data collections.
Between those, metadata can be inherited, in the sense that a piece of
metadata requested for a table is first looked up there, then in the
table's resource descriptor, and finally in the data center metadata.
This is convenient as it allows setting sensible defaults for items that
are typically constant for a data center (the publisher, say) or a
resource container (e.g., creator), while still being able to override
them in subordinate objects as necessary.

\begin{figure}
\includegraphics[width=\hsize]{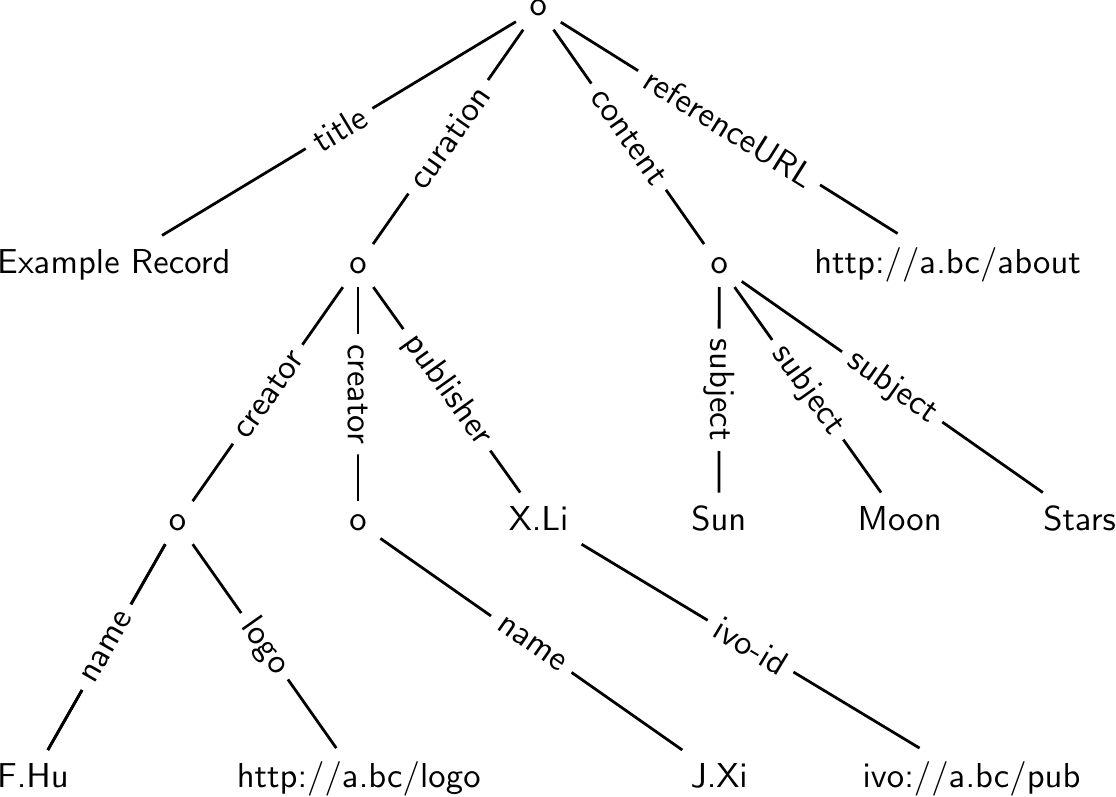}
\caption{A part of the metadata structure of a resource object within
DaCHS. 
}
\label{fig:meta}
\end{figure}

Each meta structure is a labeled tree, where each node contains a single
string, can have an unlimited number of outgoing edges, and labels on
edges need not be unique.  The labels on the edges are called
metadata atoms; concatenating the atoms
along a path with dots yields conventional metadata labels like
``curation.publisher.ivo-id'' as used, e.g., in \citet{std:RM}.

Figure~\ref{fig:meta} shows a part of such a metadata
tree containing resource metadata as defined by \citet{std:RM}. The reader may 
take a moment to appreciate the
complications of this structure, such as multiple edges with identical
labels and inner nodes with values.

This data structure can be fairly straightforwardly mapped to an XML
instance document, which is by design: Many pieces of this metadata are
used in the creation of the XML-serialised registry records.  We refrained from using full
DOM trees in the representation for several reasons.  From an
implementation perspective, our representation is more lightweight;
additional logic would be required anyway to provide metadata
inheritance; we do not want embedded fragments of VOResource XML with all
the intricacies of namespaces in DaCHS' resource descriptors; and, this
data structure is used to hold more than just VOResource metadata.

Table notes, for instance, are metadata on tables.  In this scheme, a
note consists of the note text in a node with the incoming edge labeled
note, and there needs to be exactly one outgoing edge labeled tag,
the node of which contains the note symbol as referenced in the column
definitions and used as footnote marks in rendered tables.  

This example illustrates two additional requirements that we
encountered in applications of DaCHS' metadata mechanism: Firstly, table
notes may require markup, e.g., for definition lists, enumerations, or
simply paragraphs.  Secondly, certain pieces of metadata depend on others,
as in this case a note text is meaningless without the tag.

To satisfy the first requirement, metadata content may be either
\texttt{literal}
(in which no reformatting takes place in usage), \texttt{plain} (which reflows
text in paragraphs separated by an empty line on input), and
\texttt{rst}, which
interprets the content as ReStructuredText \citep{soft:RST}, formatting
it according to the format of the embedding document.  This also
implies that in DaCHS' metadata component all nodes exclusively contain
strings.

The second requirement is covered by a primitive type system for the
nodes (rather than their content), 
which essentially constrains the children of a given
node, for instance ``\texttt{note}-typed nodes must have a \texttt{tag}
child'' or ``\texttt{news} meta items must have \texttt{author} and 
\texttt{date} children''.
Assigning types to well-known metadata items also enables 
easy input, type-specific
formatting, and defaulting where sensible.
Examples where these become relevant include \texttt{url} 
(optionally having a title
child), \texttt{relation} (having a relationship type and the related resource),
\texttt{logo} (which, formatted to HTML, yields an \emph{img}
element), \texttt{bibcode} (which, formatted to HTML, yields links to
bibliographic services), and
\texttt{info} (which are turned into VOTable INFO elements as appropriate).

A difficult design decision was whether and how to put constraints on
meta structures.  This is particularly important when meta structures
are serialised into VOResource metadata, where adherence to the
cardinality rules and content constraints implied by the schema
determines the validity of the resource records and hence the validity
of the entire OAI-PMH interface.

We opted against enforcing any constraints during metadata construction.
The most important reason is that we wanted to enable custom metadata
under operator control, as this facilitates straightforward 
extensibility as well as
flexible communication between various pieces of operator-provided artefacts
(embedded code, HTML templates, and the like).  Also incremental
creation of meta structures is much easier if unconstrained temporary
results are allowed.

On the other hand, it turned out to be highly inconvenient to diagnose
and debug bad metadata structures by schema-va\-lida\-ting the finished
resource records.  We therefore introduced a simple specification
language that constrains cardinalities of meta items. For
the service object, this might look like this:

\begin{verbatim}
title(1), creationDate(1), description(1),
subject, referenceURL(1), shortName(!)
\end{verbatim}

-- meaning that exactly one each of title, creationDate, description,
referenceURL must be present somewhere in the service's inheritance
tree, one or more of subject must be present, and, as expressed by the
exclamation point,
exactly one shortName must be in the object's own metadata.
This list is not interpreted as
exhaustive, i.e., meta keys not mentioned in this list are simply
ignored by the validator. The specifications are checked on demand,
typically by executing the \texttt{val} subcommand, or before
a resource is published to the registry.

No constraints are possible on the node content.  While this has not
been missed for simple types (e.g., text vs.~integer vs.~float),
violation of constraints imposed by controlled vocabularies has been an
issue.  As an instructive example, consider SSAP's creation type, which
has to contain one of seven keywords, some of which are long and mixed
case, like ``spectralExtraction'', as specified in \citet{std:SSAP}.
Weighing the increased complexity in the constraint language against our
experience  that vocabulary mismatches can quickly be diagnosed by
inspecting schema validator outputs, we still decided not to add
enumerations of allowed values to the constraint language.

A weak point of the validation-on-demand scheme used by DaCHS is that as
the resource descriptors are edited after publication, records may
become invalid, since resource records are generated from the descriptions
at harvest time, while the OAI-PMH timestamp remains fixed at the time
\texttt{gavo pub} was executed (which we do to avoid needlessly announcing
changed resource records to harvesting registries when
registry-irrelevant edits are made to an RD).  This can be particularly
insidious in the presence of incremental harvesting where the old,
correct record will still be present in some registries, while others
might have performed a full re-harvest in the meantime and discarded the
now malformed record.  This situation can only occur if operators fail
to re-run \texttt{gavo pub} after doing registry-relevant edits -- which
in practice happens more often than we would like.  No good solution for
this type of problem exists at that point.

\section{Resource Descriptor Techniques}
\label{sect:rd}

In DaCHS' philosophy, the
information required to publish potentially heterogeneous data over
standard protocols is collected in one XML
file, the resource descriptor.  Typically, a data center will have one 
resource descriptor per data collection.  A single DaCHS server can
expose services from arbitrarily many RDs.

As an
indicator for the extent of such descriptions, the sizes of the RDs of
services currently active in GAVO's Heidelberg data center vary between
50 lines where essentially only metadata and a service description is
necessary, and 1700 lines for resources with many tables, long notes,
and wide integration test coverage\footnote{Most resource descriptors
active in GAVO's data center are available in a public version control
system for inspection and review; a good entry point to this resource
is the cross reference at http://docs.g-vo.org/DaCHS/elemref.html.}.

Rather than discuss the mapping between the model from Fig.~\ref{fig:dcop}
and the actual XML elements and attributes in an RD -- for that, we refer
the reader to \citet{demleitner2013multiprotocol} and
\citet{soft:Dachstut} --, we want to present two case studies how the
concept of RDs aids the data collection-neutral implementation of VO
protocols.

The term ``collection-neutral'' here is crucial -- probably the single
most challenging problem in DaCHS' development and hence the evolution
of the definition of RDs has been how to enable optimal use of DaCHS'
facilities while keeping constraints on data published and
operator-defined aspects of service behaviour minimal.  Given that RDs
are written in XML, this translates into mechanisms to re-use and
customize subtrees of XML elements.  A natural choice for a technology
achieving this might seem XSLT, and if we started DaCHS from scratch,
XSLT would perhaps play an important role.  Historically, though, it
took a while until we could state the problem in this form, and so
DaCHS now offers custom facilities for this kind of metaprogramming.

One of those is the mixins dicussed above.  In the current
implementation, mixins are largely formulated using a more basic
mechanism employing \texttt{STREAM} elements (while all immediate XML
names DaCHS defines are mixed-case, metaprogramming element names are
written all upper-case), which, at the lowest level, is just a sequence
of XML parser events.  To allow customization, they support string
substitution with macros, inspired by \TeX\ (though much less
expressive).  For instance, one could define

\begin{verbatim}
<STREAM id="valWithError">
  <column name="\basename"
    description="\basedesc"
    ucd="\baseucd"/>
  <column name="err_\basename"
    description="Error in 
      \decapitalize{\basedesc}"
    ucd="stat.error;\baseucd"/>
</STREAM>
\end{verbatim}

This stream can be replayed several times in an RD, where all the macro
names not otherwise defined  must
be ``bound'' using XML attributes when replaying; this might look like

\begin{verbatim}
<FEED source="valWithError" basename="mag_v"
  baseucd="phot.mag;em.opt.V">
  <basedesc>The magnitude in the Johnson
  V band, as obtained with the ABC telescope's
  1967 V filter.</basedesc>
</FEED>
\end{verbatim}

-- \texttt{basedesc} can be written as an element since DaCHS does not
distinguish unique elements with string content and attributes.
The \texttt{decapitalize} macro used in the stream has not been bound,
as built-in macros (in this case lowercasing the first character of its
argument) ``shine through'' the bindings when replaying.

While this basic mechanism may appear rather plain, it supports a
surprising breadth of applications, in particular when combined with
\texttt{LOOP}, which inserts multiple copies of an element into an RD,
where macro bindings may be taken from a table or computed using Python
code.

An older facility still supported but not recommended for new projects
is the \texttt{original} attribute -- this references an element in some
RD, essentially turns it into a stream, and then replays it into the
element that contains \texttt{original}.  While this mimics inheritance known
from conventional object-oriented languages fairly well, it turned out
that this system had three major drawbacks in RDs. Firstly, there is a
mismatch between inheritance and the largely declarative structures in
RDs, which frequently made re-use awkward.  Secondly, we never found a
satisfying syntax for allowing changes of nodes further down the subtree
being copied.  Thirdly, and most importantly, the source objects for
\texttt{original} need to be complete, valid DaCHS objects, as they are
parsed by the normal RD parser.  Useful RD metaprogramming, in contrast,
frequently calls for node sets or incomplete objects to be moved around.
Streams can do this, \texttt{original} cannot.

While streams solved many of the metaprogramming issues we had, more
effort is still needed to regularize several important use cases, most
of which have to do with element selection; an example are
renderer-specific condition descriptors, where, for instance, an HTML
form interface will present a free-text field allowing object entry for
position selection, where a cone search needs separate RA and DEC
inputs.  To allow sharing the same core for both renderers, condition
descriptors can declare what renderer names they should or should not be
used for.  Having some general selection in DaCHS' metaprogramming
language could obviate the need for this ad-hoc construct.  However,
details of such a general selection element, in particular the language
the conditions are written in, are non-trivial to define. This results
in some rather ugly and ad-hoc DaCHS features at this point.

\section{Case Study 1: SSAP}
\label{sect:ssap}

The Simple Spectral Access Protocol \citep{std:SSAP} conceptually
is comparatively simple: There is one main operation, \emph{queryData}, with a
single-table result listing metadata for datasets matching a set of
query parameters, expressed using about a dozen protocol-specified and
arbitrarily many operator-de\-finable parameters.

However, protocol details add several complications. In addition to
responding to the simple data discovery query, the service also has to
declare metadata on the
parameters supported by it in a specific format. A wealth of metadata 
must or should be represented in the response document in either VOTable
PARAMs or the response table. Some of the protocol-defined parameters 
map poorly to many sorts of data collections (e.g., \texttt{POS} to spectra of
model atmospheres), and most parameters in turn can support some syntax
or have a relatively complex domain. As an example for the latter, we
mention the \texttt{FORMAT} query parameter, which allows specifying constraints on the
format of the datasets returned. While it could simply contain a 
MIME-type, special values like ``compliant'' -- for result
datasets serialised into one of the forms given in \citet{std:SDM} --,
``graphic'', or even ``metadata'', as well as combinations of
those, must be appropriately processed.  Finally, there are some special
validity requirements like the declaration of an XML namespace for the
fixed prefix \texttt{ssa} that is not used in the XML; this latter
requirement is due to a now-deprecated convention for data model
identifiers.

In consequence, supporting SSAP in DaCHS relies on a fairly complex
combination of definitions in
system resource descriptors -- primarily, the \texttt{//ssap} builtin
resource descriptor --
and a substantial amount of code in
the server runtime.

The first step to support SSAP was the definition of the mixins for the
tables, i.e., the columns and table parameters making up the metadata
collection.  We distinguish two cases; in mixin terms these are
\texttt{//ssap\#hcd} and \texttt{//ssap\#mixc}.  The first of these is
for publishing ``homogeneous'' collections of data, in which all datasets
originate from the same instrument and indeed the same creator.
Consequently, metadata items like the creation type, typical errors, the
bibliographic reference or the spectral resolution are table parameters
serialised into VOTable PARAM elements. The table itself only has
about 20 SSAP-related columns.

The \texttt{mixc} mixin is for a ``mixed'' data collection comprised of
spectra from multiple instruments or from multiple observing programs.
In this, information kept in table-global 
parameters within \texttt{hcd} tables now moves into each row within the table.
This roughly doubles the number of columns (and obviously the
effort to fill them from the datasets as well).  Both mixins are
accompanied by rowmaker procedures to fill the tables in a controlled
fashion; their parameter lists double as dataset metadata checklists for
the operators.

Other arrangements (e.g., constant instrument data, varying creator
data) are of course possible without code changes by simply writing an
RD deriving from the built-in \texttt{//ssap} RD.  
However, it is not clear to us yet
that the implementation effort for such mixins
is worth the moderate savings in space
and the moderate gain in data normalization.  If a consistent extraction
of constant columns into VOTable params were required, we propose this
should be done in the VOTable formatting code.

The SSAP service parameters come in the form of a stream, where the
code is written in a way that it is irrelevant
whether some piece of metadata is in the database table or
kept in a table parameter.  Implementing those service parameters
requires some extra effort, though.  About 30 lines of RD are required
for support of the \texttt{BAND} parameter, 
as it can contain a floating-point range as well as a
string to be compared against bandpass metadata.  More than 50 lines of
RD code are necessary to implement standards-compliant behaviour of
\texttt{FORMAT}.

For the remaining parameters, generic code was sufficient.  The informal
specification of what should be supported for the syntax of the
parameter values is sometimes called ``PQL'' or ``Parameter Query
Language''
and basically allows ranges and some sorts of enumerations, plus
essentially free additional metadata which we consistently ignore. It
turned out to translate into relatively involved code (about 700 lines
of Python and another 120 lines of support objects in an RD).

The built-in \texttt{//ssap} resource descriptor itself has about 800
lines pertaining to SSAP (some material in there deals with generating
datasets complying with the IVOA spectral data mo\-del).

The actual service interface again requires Python code in the form of a
DaCHS renderer.  The actual renderer can be shared with the one for the
Simple Image Access Protocol SIAP, as error behaviour and the like are
defined analogously.  The only Python code necessary at that level was
the representation of the SimpleDALRegExt \citep{std:DALREGEXT}
capability for SSAP services and about 20 lines of relatively
declarative code defining how to produce VOResource XML for that
capability.

Finally and in contrast to plain SIAP and SCS, SSAP requires a custom
core, for example in order to add the gratuitous \texttt{ssa} namespace
declaration mentioned above.  The total code necessary for the SSAP core
is some 20 lines, although it looks much more than that right now since
it is still merged with an implementation of a withdrawn proposal for a
\emph{getData} operation in SSAP.

\section{Case Study 2: Datalink}

\label{sect:datalink}

\begin{figure}[t]
\includegraphics[width=\hsize]{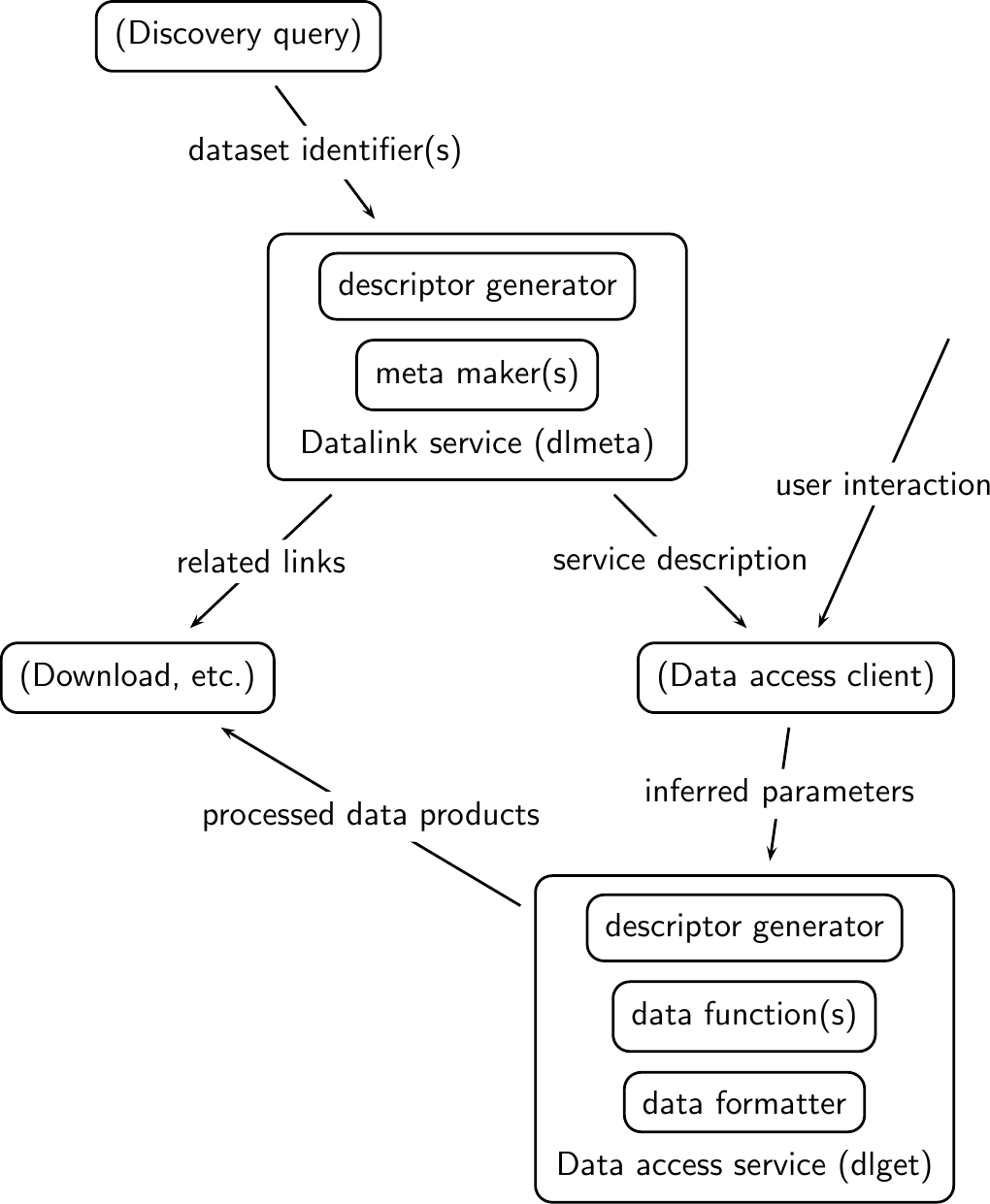}
\caption{Information flow in datalink-governed data access, with DaCHS
actors sketched in.  Items in parentheses are external.  The datalink
service proper (DaCHS dlmeta renderer) receives one or more dataset
identifiers originating from a previous discovery step.  From this, it
creates a table of links to actual data products relevant to the
dataset, as well as a description of a data access service (in DaCHS,
there is at most one of those, exposed through the dlget renderer on the
datalink service).  This allows access to processed (e.g., cutout,
rebinned, etc) versions of the dataset.}
\label{fig:datalink}
\end{figure}

Datalink \citep{std:DATALINK} is an IVOA protocol currently 
in development that abstracts
dataset delivery  by sitting between a discovery
query (e.g., using SIAP or TAP queries against Obscore tables) 
and the actual dataset delivery.  In a datalink-enabled scheme, the
discovery services may return direct links to results of datalink
services, or they may include information on what datalink services can
be used to retrieve data from the datasets discovered.  Datalink
services  themselves return 
a set of links to resources related to the dataset
(e.g., the file itself in various formats, previews, raw or further
reduced versions) as well as to data access services.

This abstraction is very beneficial to DaCHS; part of the guiding
principles behind DaCHS is to touch the inputs as little as possible,
which, before datalink, frequently meant brittle solutions to serve
standards-compliant (e.g., SDM-compliant spectra) or processed (e.g.,
cutout) datasets.  Datalink now provides a clean framework and lets
operators easily define custom operations and complex relationships in
their resource descriptors, offloading the core code.

Central to datalink is the concept of a dataset identifier (DID),
which is passed into the service to obtain the table of links
and services.  Most of the DIDs used in the VO are assigned by the
publishers, and these are known as pubDIDs for short.
PubDIDs should be IVORNs, i.e., a URI with the schema
\texttt{ivo} assigned to a specific dataset by the operator.  As IVORNs
must resolve in the registry \citep{std:VOID}, pubDIDs are typically
built as a combination of a registered resource and a query part
identifying the dataset itself.  The standard pubDIDs of DaCHS consist
of the identifier for the product deliverer, having the data center
authority as the host part and a tilde as the local part, and a query
part giving the access reference, which is a key into the product table.
This product table (containing such information as access restrictions,
physical paths, or preview location) is part of DaCHS' subsystem do deal
with possibly access-controlled files; by using such access references
in the pubDIDs, it is straightforward to re-use as much of DaCHS'
built-in data product handling as appropriate for a given task.

DaCHS' datalink implementation relies on four new procedure types.  The
first of these is a \emph{descriptor generator}, which takes a pubDID and
generates a descriptor object from it; a default implementation looks up
the DaCHS' product table and returns a bare descriptor containing
the content of the matching row, which includes information like
the physical location of the data set (e.g., a file path or a URL), the
MIME type, and the owner and embargo date for proprietary data.

Other predefined descriptor generators include one looking up SSAP
metadata and one reading the primary header of a FITS file.  Both expose
this ancillary metadata in extra attributes of the descriptor.
Generating descriptors should be fast, as descriptors are generated on
each access to the datalink service.

The next procedure type is the \emph{meta maker}.  Meta makers receive the
descriptor and come up with either parameter definitions for the
embedded data access service or link definitions which are later turned
into rows of the datalink table.  While the meta makers for specialised
data products usually presume specialised descriptors, DaCHS does not
enforce type safety here.

These two components suffice for building the datalink response.  The
\texttt{dlmeta} renderer formats the objects returned by the meta makers
into a VOTable and then delivers it to the client.

In DaCHS, datalink services typically have built-in support for the data
access services described.  For that, there is the \texttt{dlget}
renderer, which implements the data access service defined by the meta
makers returning service parameters.  Processing with this renderer
starts as for the \texttt{dlmeta} renderer but goes on to pass the
descriptor to a sequence of the third type of procedure, the data
functions.  The first of those plays a special role in that it must add a
\texttt{data} attribute to the descriptor containing some representation
of the data accessed; this could be a lazy HDU list for FITS files or a
table of spectral/flux pairs for a spectrum.

The further \emph{data functions} then perform operations defined by the
service parameters on this data, e.g., cutouts, recalibration, or
similar.  This can happen on the actual data -- DaCHS' spectral
processing does this --, but the effect can also be to add processing
instructions.  This latter option occurs for FITS files, where the data
functions handling cutouts just compute slices to be retrieved from
disk.  This lazy evaluation saves pulling large files like cubes or wide
field images into memory just to throw away most of the data.

The final step of \texttt{dlget} processing is to call a \emph{data
formatter},
the fourth procedure type for datalink support.
This takes the content of the descriptor's data item and serializes it.
The default here is trivial: The renderer simply interprets the
descriptor's data attribute as either a file or a pair of MIME type and
content (as a byte sequence) and delivers that to the user.  In the case of
spectral processing, on the other hand, the data formatter serializes
the SDM-compliant table in the data attribute into any of VOTable, FITS,
CSV, or TSV according to the value of the \texttt{FORMAT} attribute.

For special situations, data functions can shortcut the subsequent data
functions and the data formatter.  This allows the implementation of a
\texttt{KIND} parameter to the data processing service
admitting a value of \texttt{header} for FITS
files; it will inspect the current state of the data attribute and build
a FITS header out of it, which is then immediately rendered.

The implementation effort for datalink itself was rather moderate, in
particular not requiring large amounts of handling border cases.  The
core datalink code is less than 600 lines of Python, about half of which
is embedded documentation.  Code for handling special data types --
namely manipulation of spectra and generic FITS arrays -- is 800 lines
of RD including documentation.

The whole system proves very flexible.  For instance, it is
straightforward to implement delivery for simulated GAIA spectra
\citep{esa-c8spectra-report} that came in archives of the GAIA-specific
GBIN format.  For this, the only thing that needed changing with respect
to normal spectral datalink services was about 20 lines of RD describing
how to pull the spectral/flux pairs out of the database table generated
from the GBIN files.

Another example where datalink enables complex tasks without changing
the core code and still remaining compact was a prototype service
serving Echelle spectra in about 70 lines of RD.

Using the facilities, a datalink service doing standard cube cutouts is
about 20 lines of RD; by way of illustration, this is a slightly
abridged excerpt that implements such a cutout for spectral FITS cubes:

\begin{verbatim}
<service id="d" allowed="dlget,dlmeta">
  <meta name="title">Datalink service...
  <datalinkCore>
    <descriptorGenerator 
      procDef="//datalink#fits_genDesc"/>
    <metaMaker 
      procDef="//datalink#fits_makeWCSParams"/>
    <dataFunction 
      procDef="//datalink#fits_makeHDUList"/>
    <FEED source=
      "//datalink#fits_standardLambdaCutout"
      spectralAxis="1" wavelengthUnit="'nm'"/>
    <dataFunction 
      procDef="//datalink#fits_doWCSCutout"/>
    <dataFormatter 
      procDef="//datalink#fits_formatHDUs"/>
  </datalinkCore>
</service>
\end{verbatim}

A datalink service with
links to related datasets in different resolutions and a TAP service
containing the cubes in database tables is less than 50 lines of RD.

\section{Conclusion}

DaCHS is a package offering an integrated suite of tools for publishing
data with a particular focus on streamlined metadata handling from
ingestion to service operation to 
the generation of registry records.  It was written with a
view to enabling operators to adapt upstream data to standard interface
and implement custom features, always attempting to obtain
compliant-by-default behaviour.

The code is freely available under the GPL, ample documentation is
provided at http://docs.g-vo.org/DaCHS, 
and we maintain an APT repository, out of which DaCHS can be
installed to a state ready to start mapping data within minutes on
Debian and derived systems.  For
further information and downloads, please refer to http://soft.g-vo.org.

\section*{Acknowledgements}

This work was supported by BMBF grant 05A11VH3.

\section*{References}

\bibliographystyle{elsarticle-harv}
\bibliography{demleitner}

\end{document}